# Feasibility Study and Perspectives of proton Dielectric Laser Accelerators (p-DLA): from nanosource to accelerator scheme

G. Torrisi[1], D. Mascali[1*], A. Bacci[2], G. A. P. Cirrone[1], G. Cosentino[1], G. Della Valle[3], N. Gambino[4], G. S. Mauro[1], G. Petringa[1], G. Russo[5], C. Schmitzer[4], G. Sorbello[1,6], R. A. Wilhelm[7]

[1]Istituto Nazionale di Fisica Nucleare - Laboratori Nazionali del Sud (INFN-LNS), Via S. Sofia 62, 95123 Catania
[2]Istituto Nazionale di Fisica Nucleare - Sezione di Milano L.A.S.A. Via Fratelli Cervi, 201 - 20090 Segrate (Milano) – Italy
[3] Dipartimento di Fisica, Politecnico di Milano—Istituto di Fotonica e Nanotecnologie del CNR, Piazza Leonardo da Vinci 32, I-20133 Milano, Italy
[4]EBG MedAustron, Wiener Neustadt, Austria
[5]Institute of Molecular Bioimaging and Physiology, National Research Council (IBFM-CNR), Cefalù, Italy
[6]Dipartimento di Ingegneria Elettrica, Elettronica e Informatica, Università degli Studi di Catania, Viale Andrea Doria 6, 95125, Catania, Italia
[7]Institute of Applied Physics, TU Wien, Wiedner Hauptstr. 8-10/E134, 1040 Wien, Austria

**\* Correspondence:**
Corresponding Author
davidmascali@lns.infn.it

**Abstract**
In this paper we discuss the possibility to generate and accelerate proton nanobeams in fully dielectric laser-driven accelerators (p-DLAs). High gradient on-chip optical-power dielectric laser accelerators (DLAs) could represent one of the most promising way towards future miniaturized particle accelerator. A primary challenge for DLAs are small beam apertures having a size of the order of the driving laser wavelength where low charge high-repetition (or also CW) ultralow emittance nanobeams have to be transported. For electrons beams generation and acceleration, intense research activities are ongoing, and several demonstrations have been already obtained by using electrons nanotip (or flat photocathode) sources feeding dielectric microstructures. In this article we aim at the possibility to integrate a nanosource for the generation of a light ion or proton nano-beams suitable for the subsequent acceleration into sub-relativistic (low-$\beta$) p-DLA stages. Such integration includes the idea to use a proton dielectric radiofrequency quadrupole (p-DRFQ) for bridging the gap between the accelerator front-end and the drift-tube and high-$\beta$ sections. The paper has been prepared as a white book including state-of-art technologies and new solutions that now put the ambitious frontier of a fully nanostructured proton accelerator into reach. Conceptual studies of p-DLAs here presented could enable table-top proton nano-beams for several applications: proton beam writing, nuclear reaction analysis at sub-micrometer scales, the construction of miniaturized Proton–Boron Nuclear Fusion based Reactors (m-pb-NR), biological analysis at the micrometer scale, ion beam analysis at the sub-cellular level, mini-beams ion therapy to spare the shallow tissues, proton irradiation of transistors, compact proton linac for neutron generation.

**Introduction and State of the Art**

Many modern physics, industrial and medical applications require particle beams characterized by high energy up to hundred MeVs range. In this framework, Linear Accelerators (LINACs) with large accelerating gradients and ultra-compact size can be fundamental building blocks opening the way to the realization of novel devices and functionalities. Metallic LINACs operating at radiofrequency (RF) are typically used in state-of-the-art systems. Nevertheless, these structures are strongly limited by the intrinsic large losses and small breakdown threshold associated with metals, and by the lack of efficient high-power sources at high frequencies, envisaging W-band technologies which are not yet state of the art. Circular Accelerators based on Cyclotrons and Synchrotrons also require large footprint and high costs for their realization. Dielectric LINAC structures operating at higher frequencies, in particular in the infrared region, can overcome the limitations imposed by metallic structures. In particular, accelerating structures based on hollow-core dielectric waveguides exhibit larger breakdown thresholds and smaller losses with respect to metals. Furthermore, standard fabrication techniques borrowed from microelectronics allow dielectric structures with ultra-compact size to be fabricated.

An international effort – especially in the framework of the Accelerator on Chip (ACHIP) collaboration [1] - has demonstrated axial accelerating field in the GV/m range and aims to construct a "shoe-box"-size electron Dielectric Laser Accelerator (DLA) [2, 3] from below 100 keV to above 1 MeV. These efforts have led to significant technological progress over the last years, including:

- In 2013, the acceleration of relativistic electrons, firstly demonstrated at SLAC with a gradient of more than 250 MeV/m in a $SiO_2$ double grating structure driven by a 800 nm Ti: Sapphire laser [4]. In the same setup, the gradient was later increased to 690 MeV/m [5].
- Demonstrations of high gradient (850 MeV/m) and energy gain (0.3 MeV) of relativistic electrons using femtosecond laser pulses obtained in [6].
- In 2013, sub-relativistic electrons (27.7 keV) were accelerated by the group at FAU Erlangen with a gradient of 25 MeV/m using a single grating structure at the third spatial harmonic [7]. The group at Stanford University used a silicon dual pillar structure to accelerate 96 keV electrons with a gradient of more than 200 MeV/m [8] and a similar experiment at 30 keV with few-cycle laser pulses was done at FAU Erlangen [9].
- Demonstrations of optical microbunching and net acceleration of injected beams [10].
- Development of compatible laser-driven focusing for long-distance transport [11] and first demonstration of an integrated waveguide coupled accelerator on a chip [12].
- Integrated electron sources that utilize field emission from nanotips have also been developed to produce high-brightness and ultra-low-emittance (< 0.1 nm if the operating wavelength is 2 μm) electron beams extracted at 100 keV thanks to compact electrostatic lenses, well suited for coupling into optical-scale devices aiming to produce MeV-scale electron beams [13, 14].

The highest priority challenges largely concern the beam transport and matching at a given emittance in the relatively narrow (submicron-scale) apertures of DLA devices. In this context, focusing of the beam and the control of wake fields are of particular importance [15]. Recent studies of transport in extended structures with laser-driven focusing are promising for charge transport, capture efficiency, and emittance preservation [3]. According to the study in [3], ultra-low injection energies and long interaction lengths can be achieved by Alternating Phase Focusing (APF). However, many of the proposed configurations for electrons have an intrinsically limited interaction length, because they require a plane wave that impinges laterally throughout the whole structure's length. This requirement spoils the otherwise simple working principle of such interaction gratings, also known as phase-reset gratings [3]. Moreover, in the Accelerator on a Chip

International Program (ACHIP), phase stabilization and on chip waveguide laser power delivery systems have a difficult implementation [16].

**Our Proposal**

All the experiments carried out so far used electrons; the core innovation of this paper is the idea of a p-DLA along with the proton (ion)-nanosource and integration strategy. In order to have both high laser-induced accelerating gradients and adequate interaction length, it will be very helpful to use extended accelerating structures (with possibly co-linear propagation of the accelerating electromagnetic field and the particle bunch to be accelerated [17]).
In this context, photonic crystal (PhC) based structures such as hollow-core waveguides, seem to be the ideal choice for the realization of a dielectric accelerator. In fact, PhC waveguides are characterized by unique and peculiar properties:
- Hollow-core guiding structures with co-linear propagation can be easily achieved by introducing properly tailored line defects.
- The dispersive properties of the propagating modes can be carefully engineered as needed with the aim to guarantee synchronization between accelerating field and particles.
- The interaction length between particle beam and optical field is potentially rather large, in stark contrast with what happens in state-of-the-art dielectric accelerators based on laterally excited phase-reset gratings.

In order to obtain the synchrony between the travelling wave and the particle bunch to be accelerated, the waveguide phase velocity can be finely tuned, the group velocity can be engineered based on the desired interaction length and higher-order modes can be efficiently suppressed [18]. In our research plan we propose an alternative to transversely illuminated interaction structures: this consists in a collinear structure that uses the same path for the accelerating electromagnetic field and for the particle beam, in the same fashion of the standard metallic LINACs.
Preliminary results of hollow-core waveguides based on the "woodpile" photonic crystal have been obtained at scaled frequency: for example, in [19] and [20] characterizations of 3D woodpile waveguide structures, operating in the Ku and millimeter-wave band respectively, are described. In general, hollow-core structures can handle higher laser power as compared to solid core ones (for example compared to a slotted waveguide [21]), thus allowing compact high gradient acceleration stages. For this reason, dielectric structures based on hollow-core PhC waveguides are proposed in this work, since such "interaction structures" can support a suitable synchronous electromagnetic guided mode or resonant mode capable to bunch and accelerate charged particles. In particular, in presence of guided modes or resonant cavity modes, the interaction can be extended on relatively long distances in order to achieve adequate final energies.
At optical frequencies, dielectric waveguides or cavities for sub-relativistic low-$\beta$ particles are difficult to conceive and realize due to the locally dependence on $\beta$ of the geometrical features.
The use of higher harmonics or longer-wavelength (e.g., 10 µm) laser pulses, such as the demonstration experiments at the Brookhaven ATF facility [22], would enable manufacturing feasibility studies of proton acceleration with 1) higher charge throughput, 2) lower beam-emittance, 3) reduced fabrication tolerances.
The design of these low-$\beta$ sections may in principle follow the same design procedures of a conventional Radio Frequency Quadrupole (RFQ) cavity, with separate sections (cells) of functional blocks that gradually perform the bunching, focusing and acceleration of the considered particles.
The acceleration of sub-relativistic protons up to speed of light can be addressed and solved only with an integrated multi-branched solution based on the innovation of the design of the periodic interaction structure with new concepts and techniques:

- inverse design of bandgap (EBG) structures to obtain continuously matching between the phase velocity with the velocity of the accelerated particles, and an all-electric synchronous focusing system (e.g. taper slot waveguides [21]) or MEMS-based [23, 24];
- working at higher harmonics; using a reliable photonic device manufacturing platform; using phased controlled CW laser sources with proper phase control techniques (feedback loops current and cavity-length control or by direct correction of the laser frequency through external acousto-optic modulator [25]).
- tailoring and varying the shape of the interaction structure along the accelerator length in order to follow the particle energy increase with a suitable configuration for each energy regime based on hollow core 2D [17] and 3D EBG structures [19] and [20].

**Conceptual design of a new p-DLA fed by an integrated proton nanosource**

In Figure 1 we propose our new accelerator baseline including the nanosource, the injection line, the low beta and high beta sections.

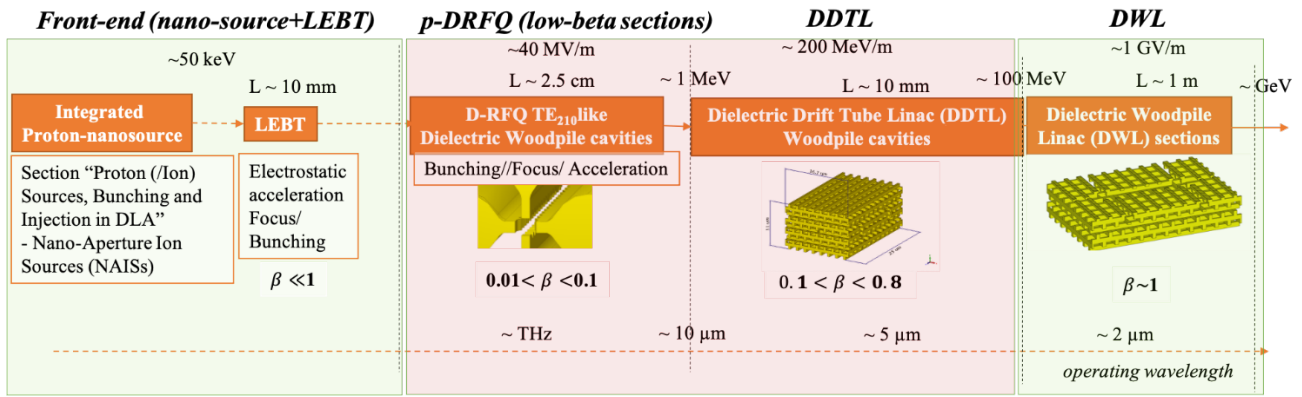

**Figure 1.** Strawman concept for proton DLA

According to the sketch of figure 1, we can conceive the DLA nanostructured baseline following the classical, energy/β based stepwise sequence of structures, starting from the proton source until the relativistic part of the accelerator (i.e., 1 GeV order of magnitude in energy, for protons).

***The front-end (nano-source+LEBT)*** of the accelerator is conceived to be based on a fully integrated proton-nanosource, whose possible technological solutions will be discussed later on in this paper (see Section "Proton (/Ion) Sources, Bunching and Injection in DLA"). In particular, block diagrams mention the one reporting the most suitable characteristics for a proper matching and integration in the accelerator front end: i.e., the "NanoAperture Ion Sources" (NAISs). The front-end is then thought to continue with a compact electrostatic part able to accelerate the beams up to approximately 50 keV (this value could be increased at expense of longer extraction gap distance or greater extraction field). This energy – along with the extreme compactness of the source-beamline ensemble – is already of interest for some relevant biomedical studies that will be discussed below (see Section "Envisaged Applications").

***Proton Dielectric Radio Frequency Quadrupole-DRFQ (low-beta sections up to 1 MeV)***
The device proposed to bridge the gap towards intermediate and high (relativistic) energies is hereby a dielectric RFQ-equivalent system. For the first time this is proposed in a p-DLA chain, and very preliminary simulations were carried out in order to verify the feasibility of such accelerating structure in terms of proper modal distributions for the beam focusing, in addition to the acceleration scheme. The main advantage of a nanostructured high-frequency dielectric RFQ-like section is the possibility to reach, like for the other sections, much higher field gradients in order to miniaturize the system.

***Dielectric Drift Tube Linac (DDTL, up to 100 MeV)***: this part is also inspired by the well-known structures in metallic accelerators named Drift Tube Linac. Woodpile technology can be employed to obtain a sequence of different aspect ratio cavities in order to match the phase velocity to the variable particle velocity [29]

***Dielectric Woodpile Linac (DWL at relativistic scale, up to1 GeV)***
Results are reported in [20, 30] concerning electrons, but simulations are ongoing for p-DDTL conceptual design.

With this paper being focused on protons acceleration, we would investigate more in detail the issues of the low-energy section of the accelerator.

In laser-driven devices, the energy gain per optical period is very small (typically on the order of a few keV, at most) and the bunch will have $\beta=v/c$ well below unity for tens or hundreds of periods. Any DLA concept must therefore include a method for sub-relativistic pre-acceleration: both the particle velocity and the structure parameters must also change to maintain continuous acceleration over that velocity range (for example, from 0.3c to 0.8c). Being the synchronicity condition expressed as $\beta=\Lambda/m\lambda$, where $\beta$ is normalized (to c) velocity, $\Lambda$ is the periodicity of the accelerating structure, *m* the harmonic number and $\lambda$ is wavelength, the spatial periodicity of the hollow-core channel needs to be adjusted along the accelerating structure in order to allow the particle bunch to gain energy from the synchronous accelerating mode.
The most significant challenge for a successful injection is the mismatch in energy between a candidate particle sources and the near-relativistic energy regime in which DLAs operate most effectively. Frequency jumps in an ion LINAC can be used in order to provide a large transverse acceptance in the low-energy part and a high accelerating gradient in the high-energy part [26]. A smooth tailoring of the dispersion properties of photonic crystals guiding structures allows to manage the frequency transition, reducing the discontinuity which could potentially affect the longitudinal acceptance of the LINAC if not carefully addressed. For each frequency regime where dielectric accelerating structures represent a viable solution, we inserted the main references and features in the Table 1 (except the first row where we included metallic W-wave accelerator as a boundary layer between the metallic and the dielectric "world"). In filling the table, we considered not only the intrinsic physical mechanisms of the ions acceleration, implying the above mentioned needed frequency jump, but also the feasibility of each section realization, at a given frequency/wavelength domain, in terms of pumping sources availability and fabrication technology. Proposed solutions are based on wavelength scales compliant with well-established technology readiness. However, operating wavelength numbers reported in the table have to be considered as indicative of the needed different regimes, whilst in the future implementation of the accelerator feeding systems a smooth progression of the operating frequency has to be taken into account.

Table 1: Main refs. For each frequency regime

| Operating Wavelength | Pumping Source | Manufacturing | Ref. | material |
|---|---|---|---|---|
| mm-wave | Gyrotron | conventional machining techniques | [27] | metal |
| THz | THz Gap | conventional machining techniques | [28] | **dielectric**-loaded circular waveguide |
| ~10 µm | $CO_2$ Laser | sub-micron precision | [22] | **dielectric** |
| **~1-2 µm** | **Solid-state Laser** | **Nanometer precision** | **[3]** | dielectric |

Moreover, at the micron scale, there is no equivalent to the conventional RF gun, in which electrons are brought to relativistic energies in just one or two wavelengths.

It should be noticed that the reduction of the minimum velocity of particles that can be accelerated is essential in order to miniaturize the overall particle source and to relax the requirements of the pre-acceleration stage. Sub-relativistic or "low-beta" electron DLA-like structures - capable of optically accelerating electrons from tens of keV to roughly 1 MeV - can be created using periodic material patterns (shapes) near the beam path through non-resonant, nearfield structures in which the optical phase can be controlled to a high level by detailed adjustment of the structure periodicity and coupling. In the case at the hand, the development of low-beta acceleration schemes for proton and ion acceleration, could be done by making portions of the beam path nearly field free, thus shielding the beam from off-phase deceleration, as is in a drift-tube ion linac.

In resonant or standing-wave devices asynchronous scheme could be used, in which a mixing of optical wavelengths leads to net acceleration, even though the particle and phase velocities are never synchronous. This scheme requires a subharmonic periodicity either with multiple-wavelength illumination or with monochromatic illumination of a subwavelength structure.

Low-β accelerating sections can be obtained either by decreasing the structural periodicity of the HODAC, by increasing the order of the harmonic [2, 7], or by increasing the wavelength of the source. The minimum value of periodicity is limited by the fabrication process. As a consequence, at higher frequency it may be convenient to use higher spatial harmonics to accelerate sub-relativistic protons.

The strategy of a low-β accelerator stage has to consider the bunching and acceleration of low-β. The first sections of low-β particle acceleration can also be achieved by using a higher-order TE-mode (i.e. a quadrupole mode) capable of gradually bunching and accelerating the particles. Such TE-like structures use a transverse electric field to focus and a longitudinal field to accelerate the particles, as in the conventional RFQ structure. The axial field component is generated by a proper cavity modulation, in order to create cells with increasing length as the particle bunch travels along the structure. Focusing, bunching and acceleration of low-velocity particles depend on such modulation and will be optimized by automating the design process through inverse design technique, which allow the desired properties to be obtained by means of local perturbation of a TE-like PhC Cavity.

Beam dynamics (BD) of heavy particle acceleration (protons or ions), compared with electrons, cope with complex technological aspects of the accelerator's devices. The main motivation is that heavy particles need a higher kinetic energy to reach the speed of light, therefore acceleration at low-β becomes crucial. Let us consider a low performing DC electron gun extracting an e-beam at 100 keV, by the relativistic relation with βγ · √(γ^2-1)/γ with γ =1+T/ E0 (E0 the rest mass energy) β results to be 0.548, while for 100 keV energy protons it is 0.014 (~40 times lower). A main acceleration condition is to set the acceleration wave phase velocity equal to the beam velocity, which implies acceleration gap length equal to βλ. For non-relativistic beams, with a quick beam energy variation, the acceleration gap length (or the cavity length) needs to change continuously.

For electrons this concern is rarely stressed, because electrons are relativistic in short accelerating gaps and at the energy of few hundreds of keV just slips few degrees on the acceleration wave, an effect that usually can be controlled by tuning the injection phase.

Heavy particles are also characterized by a high rigidity Bρ=p/q. For protons Bρ scales linearly with the mass, accordingly the focal capability of quadrupoles scales in the same way.

Low energy electron beams can be fully controlled with short (cm's orders) and weak quadrupoles gradient e.g., at energies of 0.1- 0.5 MeV with less than 0.1 T/m, when protons in the same energy range need a gradient of about 200 T/m. Few years ago these values would have been too high; nowadays with permanent magnet quadrupoles is possible to reach gradient up than 500 T/m [23]. Because the gradient scales with the quadrupole bore size and because DLA accelerators work with narrow beams (submicron-scale), these gradients could be also higher: 500 T/m considers mm bore size, further exist other promising technologies as the micrometrical systems (MEMS) [24]. In order to deal with the necessary compensation of the emittance pressure, we will need to design a proper integrated lattice of these devices. We are well aware that the emittance requirements represent an issue where a solution is still missing, and this could represent a showstopper for p-DLA. Current limits and criticalities for electrons have been recently shown in [11]. Even more challenging will be for protons. Considering for example a proton beam coming from a Gas filled Ion Sources (GFIS), we assume at the DLA injection the following parameters: E = 50 keV, ΔE/E ~ 2x10-5, εn= 10-9 rad, a spot size of ~200 nm. During the acceleration the beam will be bunched into the acceleration buckets, that, considering for example an acceleration frequency of ~200 THz, (1.5 μm) results into a bucket length of~ $c/f \cdot \beta$ ~15 nm, $\beta$ in the order of 0.01. A so short bunching, from space-charge point of view, is possible because of the average beam current order of 10 pA, (a low charge per bucket), indeed low-β protons bunches suffer of emittance degradation by space-charge which contribute scales as $1/\gamma^3$. Because of the nano scale size technologies challenging at ~50 keV it seems favorable the use of a D-RFQ. At energy order of ~1 MeV ($\beta$~0.05) few accelerating constant length buckets (e.g., a wood-pile DLA) show an effective acceleration also considering a strong bunch/phase slippage, that means that a DLA staging, with constant size per device can be considered; at energy higher that ~100 MeV $\beta$~0.5) the staging could be done with quite longer cavities and effective acceleration.

**Proton (/Ion) Sources, Bunching and Injection in p-DLA**

All accelerators require the creation and injection of a particle beam with properties (size, time structure, and energy) that must match the dimensions and fields of the high-energy accelerating structure. Promising candidate mechanisms for microscale electron sources, and techniques for bunching them, are discussed in [2, 14]. Furthermore, the problem to match an electron beam into micrometre size acceleration buckets is under study since years in electron plasma wave accelerators field; very promising results has been already carried out [31]

For integrated proton nanosources for DLAs we can start from [32] analysing ion sources based of different mechanisms (Electron impact (EI), chemical ionisation (CI), fast atom bombardment (FAB), field Ionisation and field desorption (FI/FD), matrix-assisted laser desorption (MALDI), atmospheric chemical ionisation, electrospray ionisation (ESI)).

On the basis of the considerations reported in the previous Section "Our Proposal: Conceptual design of a new p-DLA fed by an integrated proton nanosource" about the low-beta sections of the p-DLA, in Table 2 we listed nanostructured proton/ion sources requirements. Energies of about 47 keV ($\beta$=0.01) are required with a beam size on the order of 100 nm of minimum diameter, sub-mrad divergence, ultra-low emittance (nm-rad).

**Table 2: Preliminary Estimated Source Requirements**

| Parameter | Value |
|---|---|
| Proton Energy | ~ 50 keV ($\beta$=0.01) |
| Diameter | << 1 µm |
| Emittance | nm-rad |
| Beam format | CW or Pulsed @high rep. rate laser synchronized |

In the following, nanostructured proton/ion sources possible solutions have been preliminarily identified and described:

- **Gas filled Ion Sources** (GFIS) [33-35]: one of the possible solutions suitable for p-DLA are the Electron Ionisation (EI) sources - where vapour-phase molecules are ionised by energetic electrons (~70 eV) that strip a valence electron from the analyte, resulting in the formation of an $M^+$ molecular ion with concomitant fragmentation - and Field Ionisation. Electron-impact ionization theory has established the following relationship for the generated ion current $I_i = kI_e l\rho\sigma(E)$ where k is a device specific geometrical constant, $I_e$ is the electron current, l is the electron path length, $\rho$ is the neutral gas density, and $\sigma(E)$ is the ionization cross section which depends on gas type and electron energy. Electron-impact ionization cross sections for most atoms and molecules have maximal levels typically between 70 and 100 eV.

    In Field Ionisation sources, molecules passing through a very high electric field are ionised by removal of an electron via quantum tunnelling to produce an $M^+$ molecular ion. The gas filled ion source is a technology that has been commercialized over the past decades as the ion source for helium ion microscope (HIM). GFIS has been mostly used for generating He and other noble gases such as Ne ions, but they can be extended to other gaseous elements such as hydrogen. The HIM and the GFIS technologies have some very impressive properties, such as a sub-0.5-nm probe size, high brightness (4.00 GA/cm² sr), low energy spread ($\Delta E/E \sim 2 \times 10^{-5}$), and an image resolution of 0.35 nm for lithography applications. These numbers translate to an emittance of roughly $1 \times 10^{-11}$ m rad [36]. The ions are generated through a high electric field above a single atom in the presence of a noble gas. Field ion microscope in its simplest form consists of a cryogenically cooled tip, biased to a high voltage. The strong electric field is concentrated on a super sharp metal needle, in general the apex of a pyramidal tungsten tip.

When a high voltage V is applied to the tip, the generated electric field E can be approximated by E~ V/5r wherein r is the radius of curvature of the tip. The applied voltages range up to 20-25 kV, higher extraction voltages tend to increase the number emitting regions on the tip, degrading the angular current density and depleting the gas supply. Reported beam currents range from 1 fA up to 100 pA [35] but are usually optimized for minimum virtual source sizes and not for maximized beam current. Such nanosources are typically used for helium ion scanning microscopes (HISM) which use a GFIS in which a tip supplied at high voltage and low temperature to ionize helium gas.

- **Electron-impact gas ion sources (EIGIS)**: direct ionization by "electron-ion impact" of an $H_2$-jet, with a suitable source of cold cathode electrons (for reasons of reliability, duration, purity, etc.). [37]: the device consists of a cathode with aligned carbon nanotubes, a control grid, and an ion collector electrode. It is suitable for direct integration [38, 39]. Ion currents in excess of 1 µA can be generated. In [39], a cold cathode electron impact ion source made from an array of diamond-coated silicon whiskers for application in an ion trap mass spectrometer, is described.

CNTs can have average diameter of tens on nm. These devices have cathode-to-grid spacing of tens of µm and a grid-to-collector spacing of hundreds of µm. The cathode is a panel with a matrix of CNT. The characterization of the electron-impact ion source can be carried out at room temperature in a vacuum chamber with a base pressure of $10^{-6}$ Torr.

The electric fields required to generate 1 nA and 1 µA of electron current are 5 and 6 µV/m, respectively. These devices were routinely capable of generating field emitted electron current in excess of 50 µA. Of course, the use of the electron beam requires an appropriate tuning of the energies to maximize the ionization rate, and this depends on the ionic species. The paper [40] indicates that about 100 eV energy of the electrons is sufficient to have the maximum ionization rate.

  - A special case of EIGIS are the so-called **Nano-Aperture Ion Sources (NAIS)**: the conversion of an electron beam into an ion beam works by electron impact ionization (see above). Using a high brightness electron source as for example a Shottky emitter type source or a thermal $LaB_6$ emitter in a SEM allows the focussing of an electron beam into a small volume with high current densities [43, 44]. At optimal conditions of a low energy (10-100 eV) electron beam, current density of ~$10^{22}$ e/cm²s can be reached and thus a conversion efficiency of 100% is possible, i.e. every electron facilitates an ionization event in the anticipated ionization volume provided a reasonable amount of atoms in the gas phase. In [47, 48] which most likely would be interesting as the ideal solution: these ion sources have been developed for proton direct beam writing (or Direct Write Lithography (DWL)) and other focused ion beam applications [45, 46]. The latter demands the use of a gas jet or a sophisticated differential pumping system in order to maintain high vacuum conditions outside the ionization volume mitigating electron and ion scattering. With typical aperture sizes of 100 nm (some are discussed with 1 µm or larger) and aperture spacing of 500 nm, the ionization volume amounts to about $4 \times 10^6$ nm³. A gas pressure of about 100 mbar is necessary to ensure on average at least one gas atom/molecule is in the small ionization volume. Design studies use about 1 bar gas pressure in the ionization cell [45, 46]. Increasing the ionization volume relaxes this requirement but diminishes the brightness. Realizing a NAIS in a compact way would most likely be the ideal solution for a DLA injector. A NAIS is supposed to deliver sub-10 nm beams with a current of 0.5-1 nA. The ionization cell itself can be manufactured as an integrated device, thus even in combination with the lithography necessary for the DLA. In this way, no alignment issues will arise. Using a suitably high extraction potential, energies of 30-50 keV can be reached, meeting the injection requirements of a DLA together with a small beam divergence of <5 mrad.

- **Laser-based sources**: utilizing a state-of-the-art infrared laser source (800-1000 nm) with pulse durations of less than 100 fs and energies per pulse of about 1 mJ yields power densities of $10^{14}$ W/cm$^2$ and higher upon proper focusing in a gas volume or gas jet. With these power densities multi-photon ionization with a probability of up to 100% can be reached to produce ions close to the entrance of the DLA [41]. Applying a strong electric field at the same time at the ionization volume in the range of 5 kV/mm or higher enables acceleration of the ions to the minimal acceptance energy of the DLA over a distance of about 10 mm. However, these high intensity lasers are typically limited to repetition rates of about 1 kHz and this frequency is mostly fixed. Another possibility is the use of a low power laser system with 0.1-5 µJ per pulse and a higher repetition rate of 20-100 kHz at about 500 nm (second harmonic), anyway too low for pDLA applications. While multi-photon ionization is not possible with the low pulse powers, the pulses can be used to trigger ionization from a field emitter tip as it is done in fs-laser assisted atom probe tomography [42] or potentially in field ion microscopes. Utilizing the field enhancement at the tip apex enables the ionization of a surface atom or the tunneling of an electron from a gas atom in the vicinity into the tip, respectively. The resulting ion is accelerated from the tip due to a DC electric field. This approach is beneficial with respect to the resulting beam emittance since the ionization volume is in the nm$^3$ range. One major drawback result from unavoidable laser power fluctuations which may cause irreversible tip damage and consequently hinders the [long-term operation as an ion source. It is, however, an engineering problem and not a general physical limitation implying some hope to overcome these issues in the future.
- **dielectric capillaries** to guide and focus the microbeams [47-50].

**Dedicated Beam Diagnostics**

The diagnostics of a sub micrometric proton beam represents an extreme challenge if a suitable spatial resolution has to be ensured, in order to measure the size of the transversal profile with a reasonable uncertainty. A performing and reliable device needs to adopt a state-of-the-art approach, based on innovative technical solutions to overcome the current limits in spatial resolution that hardly are lower than a few micrometers. Two possible approaches to be followed for our purposes can be taken into account as possible candidates, in order to ensure a spatial resolution of 200-300 nm for proton beam intensity of a few pA. One of them is the optical approach, which can be sketched with an optical microscope coupled with a CCD camera, watching an extremely thin scintillating screen crossed by the proton beam. The energy lost by the protons in the screen is converted in a visible light spot whose size is given by the convolution of the true beam size with the light dispersion inside the screen itself. A further contribution to the resolving power is due by the spatial resolution obtainable with a microscope that can reach at most 200 nm. Moreover, the thickness of the screen has to be chosen as a careful tradeoff to maintain a low light dispersion with a thin layer, but at the same time guaranteeing the needed light intensity. For such a reason a high efficiency scintillator represents the proper choice and the CsI(Tl) could be a possible good candidate, having a very high efficiency of about $55 \cdot 10^3$ ph/MeV and a wavelength of the scintillating light that well matches with most of the standard CCD camera. The sensitivity can be further improved by adopting high sensitivity CCD camera among the several available models, with cooled CCD and with the option of long-time exposure to reduce the statistical noise. In [51, 52] light spots of proton and ion beams with intensity down to hundreds of fA and energy tens keV or less have been obtained with CsI(Tl) screens, whose the epitaxial deposition of hundreds nanometers on a substrate can be obtained. The other possible approach is based on thin metallic strips working as a scanning wire system, that can miniaturize at on chip level and meet the requirements of resolution and sensitivity. In [53] beam profiles of 500 nm have been measured with a spatial resolution of 250 nm, by exploiting sub micrometer Au metallic strips scanning the transversal beam plane. The work is very promising and the adoption of SiC strip could further improve the precision of a factor 3. With the perspective of a direct matching with the on-chip proton source and p-DLA it seems to represent the valid solution for portable systems. Similarly, the metal wires could be used as a target for Particle Induced X-Ray Emission (PIXE), where characteristic X-Rays from the metal constituents are registered in a separate (non-spatially resolved) detector only when the nano-ion beam impacts the wire.

**Envisaged Applications (Medical, cell biology)**
The added value to go beyond electron acceleration, with proton nanobeams, could open the possibility to plenty of applications, from medicine to nuclear fusion such as proton beam writing [45], nuclear reaction analysis at sub-micrometre scale [54], Proton–Boron Nuclear Fusion Reaction [55], biological analysis at the micrometre scale [56, 57], ion beam analysis at the sub-cellular level [58, 59], mini-beams ion therapy [60, 61] to spare the shallow tissues.
Light ion radiotherapy became in the last decade an established technique for cancer treatment. Currently 13 operational centers worldwide offer carbon ion therapy and only six are dual particle therapy facilities combining protons and carbon ions [62]. As mentioned in the introduction, large footprint accelerators based on cyclotrons and synchrotrons are used for the above-mentioned applications. Proton therapy centers use energies between 62.4 and 252.7 MeV, corresponding to beam ranges from 30.0 to 380.0 mm in water. The proton spot sizes (applicable for pencil beam scanning) are on the order of 10-20 mm at minimum energy and of 5-10 mm at maximum energy. The required beam intensity relates to the total number of particles filling the main accelerator over the extraction time. In the case of the synchrotron-based accelerator MedAustron based in Austria, this intensity corresponds to $2\cdot10^{10}$ p/s. The extraction time (i.e. the length of a required spill length) lies in a range between 1 and 10 s. Light ion beams can be delivered using different approaches, including the passive scattering technique, uniform scanning, and pencil beam scanning (spot scanning or raster scanning). The required field sizes for tumor scanning are on the order 20·20 cm at the nominal beam delivery position. The beam delivery is generally a controlled Dose Delivery System [63]. While being able to deliver energies that can be used to cover common tumor sizes and locations the accelerator has to deliver also an acceptable dose rate for treatment, which is should be at least on the order of 2 Gy/min. Moreover, the efficiency of a beam delivery system is about 20% because of scattering effects.

The development of radiation microbeams has been and still is an exceptional tool for the investigation of the fundamental mechanisms staying at the basis of cell biology response to the radiations. The power of micro-irradiation approaches lies in the capacity to deliver very accurate doses to individual and well-identified cells in vitro. The possibility to irradiate sub-cellular structures is, of course, also possible.
All these potentialities have led to the growth of a number of worldwide microbeam facilities. These allow for the delivery of precisely defined beams of charged particles, X-rays, or electrons. Given their precision, microbeams can act as probes to fully understand all the events that stay, for example, below the DNA damage response after a localized dose deposition.

In particular, microbeam' facilities for cellular response research require the use of sophisticated high throughput technologies. On the one hand, it is necessary to use technologies that make it possible to precisely deposit few cells on a support. This is an important point, as the cell positioning need to be accurate, precise and reproducible, by using automated dispensers. On the other hand, advanced microscopy technologies are also necessary, both to target cells or its sub-cellular organelles, to hit by irradiation, both to follow the outcome of irradiation through a live-imaging approach. The best technological solutions are integrated systems, consisting of microplate stacker including dispensers, detectors and imaging systems. The last ones, could be fluorescent microscopes, which are very helpful in following the labelled key protein through the use of Fluorescent protein-tag (GFP-Tag). This approach, offers the opportunity to study the spatio-temporal involvement of DNA repair factors, to deeply explore mechanisms induced by low doses of ions or particles, especially, if the microscope is directly connected to a cell incubator, permitting to perform live-imaging over hours and days after irradiation, under controlled temperature and $CO_2$ conditions. However, other label free imaging techniques could be used, such as quantitative phase microscopy or phase contrast microscopy.

In a review, Price and Schettino [64] discuss the radiobiologic advances in this research field. They described different experiments performed with microbeam irradiation of cells and tissues. Among the experiments described, it was clearly highlighted the bystander effect, observed on the non-directly irradiated cells and tissues areas. This finding is also observed even irradiating exclusively cytoplasm, which is activated by water radiolysis, generating ROS [65]. For example, Shao et al [66] observed bystander responses in radioresistant glioma cells, even when only the cell cytoplasm is irradiated. The opportunity offered by the solely cytoplasm irradiation demonstrates the fact that DNA is not the exclusive key target of ionizing radiation in generating biologic effects and clarify the cell-cell communication role.

However, the microbeam irradiation and live imaging technology offer another important chance of radiobiology research, in exploring the kinetic and localization of DNA repair involved molecules. This is a great breakthrough, as in the classical in vitro cell irradiation, the DNA damage can be evaluated, by the classical $\gamma$-H2AX foci or some other labelled key molecules appearance, on fixed cells.

By a live imaging approach and microbeam irradiation, it is possible, for example, to better understand the Double strand break (DSB) repair steps, i.e how chromosome ends from different DSBs meet each other to form chromosome aberrations ([64]). In addition, by this approach it is possible to follow the destiny of some proteins of interest, elucidating their role in the DNA repair pathway. This approach was recently used by Jacob et al [67], to assess that DSB repair kinetics and fidelity are dependent on the damage entity.

Finally, it has to be cited the opportunity deriving from microbeam irradiation of 3D cell culture or thin tissues, very useful, once again, in revealing the bystander effects in the close cells. The first experiments on tissues were performed on human and porcine ureters, which have the characteristic to be organized with four to five urothelium layers. The microbeam offers the possibility to hit just single small section of four to eight urothelial cells [64]. In addition, commercially available skin models have been also used by Belyakov and co-authors[68], which observed micronucleated and apoptotic bystander cells up to 1 mm away from irradiated area, suggesting a prominent role of cell-to-cell communication by gap junctions and of autocrine and paracrine molecules.

Further advances in very precise beam delivery have also enabled the transition towards new and exciting therapeutic modalities developed at synchrotrons to deliver radiotherapy using plane parallel microbeams, in Microbeam Radiotherapy (MRT).

The chance to have controlled and reliable nanosource of proton and ion beams, with energies ranging from 50 keV (for protons, corresponding to a stopping power in water of 90 keV/um) to 50 AkeV and 15 AkeV for Helium and Carbon (corresponding to stopping power in water up to 200 keV/um), respectively, will lead to the unprecedented possibility of investigate the cell-damage mechanisms in a deterministic way and at the DNA level

**Conclusions and Perspectives**

In summary, the proposal of a fully DLA for protons presented in this paper aims s to work at optical frequencies with dielectrics and develop novel interaction structures providing a continuous synchronicity and acceleration for protons from sub relativistic to relativistic regimes. Such miniaturization will provide unprecedented capability in charged particle acceleration and place the basis for a roadmap of proton nanoaccelerators development with impact at each energy range, including sub-relativistic ones at 50 keV until the relativistic scale at 1 GeV. In order to mitigate current limitations due to the reduced channel cross section an option is to operate with a Continuous wave (CW) electromagnetic field, and we could also adopt a multichannel device which forms a DLAs matrix [69].

As an ultimate and long-term goal, it can be said that such miniaturization could contribute to develop much more compact cancer therapy accelerators, as well as other applications such as p-B ultracompact nuclear fusion devices based on proton acceleration below 1 MeV. At the current state of the art, the current medical facilities require synchrotron-based accelerators which occupies large areas in the size of an entire building. Moreover, conventional accelerators require also large beam delivery systems based on complex magnetic systems which allow to move the beam within the needed scanning area [63, 70]. An array of DLAs could be potentially used to irradiate the tumor area with the accelerator with a proper shaping of the beam or, even more interesting, due to the extreme compactness of the p-DLA scheme, the accelerator itself is suitable to be quickly moved in order to perform a fast scan of the tumor mass to be irradiated.

As next steps for the continuation of the R&D we will focus on the design of the p-DRFQ and the beam dynamics across the entire accelerating chain. THz and laser sources together with dielectric couplers between supplying sources and accelerating channels have been already investigated for the high energy sections (woodpiles), as mentioned above, but will be also extended to the sub-relativistic section for future studies on such high frontiers accelerating structures.

**Acknowledgements**

R.A.W. acknowledges funding for the Austrian FWF, project no. Y1174-N36.

**Reference**

[1] ACHIP website: https://achip.stanford.edu

[2] R. Joel England et al. Dielectric laser accelerators. Rev. Mod. Phys. 86, 1337 (2014)

[3] U. Niedermayer, et al; "Challenges in simulating beam dynamics of dielectric laser acceleration" International Journal of Modern Physics A Vol. 34, No. 36, 1942031(2019) https://doi.org/10.1142/S0217751X19420314

[4] Peralta, E., Soong, K., England, R. et al. "Demonstration of electron acceleration in a laser-driven dielectric microstructure". Nature 503, 91–94 (2013). https://doi.org/10.1038/nature12664

[5] K. Wootton, Z. Wu, B. Cowan, A. Hanuka, I. Makasyuk, E. Peralta, K. Soong, R. Byer, and R. Joel England, "Demonstration of acceleration of relativistic electrons at a dielectric microstructure using femtosecond laser pulses," Opt. Lett. 41, 2696-2699 (2016).

[6] Cesar, D., Custodio, S., Maxson, J. et al. High-field nonlinear optical response and phase control in a dielectric laser accelerator. Commun Phys 1, 46 (2018). https://doi.org/10.1038/s42005-018-0047-y

[7] J. Breuer and P. Hommelhoff, "Laser-Based Acceleration of Nonrelativistic Electrons at a Dielectric Structure"" Phys. Rev. Lett. 111, 134803 – Published 27 September 2013

[8] Kenneth J. Leedle, Andrew Ceballos, Huiyang Deng, Olav Solgaard, R. Fabian Pease, Robert L. Byer, and James S. Harris, "Dielectric laser acceleration of sub-100 keV electrons with silicon dual-pillar grating structures," Opt. Lett. 40, 4344-4347 (2015)

[9] Kozak, M., "Dielectric laser acceleration of sub-relativistic electrons by few-cycle laser pulses" Nuclear Instruments and Methods in Physics Research A (2016), http://dx.doi.org/10.1016/j.nima.2016.12.051

[10] Dylan S. Black, Uwe Niedermayer, Yu Miao, Zhexin Zhao, Olav Solgaard, Robert L. Byer, and Kenneth J. Leedle "Net Acceleration and Direct Measurement of Attosecond Electron Pulses in a Silicon Dielectric Laser Accelerator", Phys. Rev. Lett. 123, 264802 – Published 26 December 2019

[11] Uwe Niedermayer, Thilo Egenolf, Oliver Boine-Frankenheim, and Peter Hommelhoff, "Alternating-Phase Focusing for Dielectric-Laser Acceleration" Phys. Rev. Lett. **121**, 214801 – Published 20 November 2018

[12] Sapra et al., Science 367, 79-83 (2020)

[13] A. C. Ceballos. Silicon Photocathodes for Dielectric Laser Accelerators. PhD thesis, Stanford University, 2019,

[14] T. Hirano et al; "A compact electron source for the dielectric laser accelerator", Appl. Phys. Lett. 116, 161106 (2020); https://doi.org/10.1063/5.0003575

[15] B. Naranjo, A. Valloni, S. Putterman, and J. B. Rosenzweig "Stable Charged-Particle Acceleration and Focusing in a Laser Accelerator Using Spatial Harmonics" Phys. Rev. Lett. 109, 164803 – Published 19 October 2012

[16] Tyler W. Hughes, Si Tan, Zhexin Zhao, Neil V. Sapra, Kenneth J. Leedle, Huiyang Deng, Yu Miao, Dylan S. Black, Olav Solgaard, James S. Harris, Jelena Vuckovic, Robert L. Byer, Shanhui Fan, R. Joel England, Yun Jo Lee, and Minghao Qi, "On-Chip Laser-Power Delivery System for Dielectric Laser Accelerators", Phys. Rev. Applied **9**, 054017 – Published 14 May 2018

[17] G. Torrisi et al 2019 J. Phys.: Conf. Ser. 1350 012060

[18] J. M. Fini, "Suppression of higher-order modes in aircore microstructure fiber designs," 2006 Conference on Lasers and Electro-Optics and 2006 Quantum Electronics and Laser Science Conference, Long Beach, CA, USA, 2006, pp. 1-2, doi: 10.1109/CLEO.2006.4627817

[19] G. Torrisi et al., "Design and Characterization of a Silicon W-Band Woodpile Photonic Crystal Waveguide," in IEEE Microwave and Wireless Components Letters, vol. 30, no. 4, pp. 347-350, April 2020, doi: 10.1109/LMWC.2020.2972743.

[20] G. S. Mauro et al., "Fabrication and Characterization of Woodpile Waveguides for Microwave Injection in Ion Sources," in IEEE Transactions on Microwave Theory and Techniques, vol. 68, no. 5, pp. 1621-1626, May 2020, doi: 10.1109/TMTT.2020.2969395.

[21] Zhexin Zhao, Tyler W. Hughes, Si Tan, Huiyang Deng, Neil Sapra, R. Joel England, Jelena Vuckovic, James S. Harris, Robert L. Byer, and Shanhui Fan, "Design of a tapered slot waveguide dielectric laser accelerator for sub-relativistic electrons," Opt. Express 26, 22801-22815 (2018)

[22] W. D. Kimura, I. V. Poaorelsky and L. Schächter, "$CO_2$-Laser-Driven Dielectric Laser Accelerator," 2018 IEEE Advanced Accelerator Concepts Workshop (AAC), Breckenridge, CO, USA, 2018, pp. 1-5, doi: 10.1109/AAC.2018.8659403.

[23] J. Harrison, Y. Hwang, O. Paydar, J. Wu, E. Threlkeld, J. Rosenzweig, P. Musumeci, and R. Candler Phys. Rev. ST Accel. Beams 18, 023501 – Published 17 February 201

[24] Jere Harrison, et al, "High-gradient microelectromechanical system quadrupole electromagnets for particle beam focusing and steering", Phys. Rev. ST Accel. Beams 18, 023501 – Published 17 February 2015

[25] T. Sala, D. Gatti, A. Gambetta, N. Coluccelli, G. Galzerano, P. Laporta, and M. Marangoni, "Wide-bandwidth phase lock between a CW laser and a frequency comb based on a feed-forward configuration," Opt. Lett. 37, 2592-2594 (2012) https://www.osapublishing.org/ol/abstract.cfm?URI=ol-37-13-2592

[26] Frequency jump in an ion linac R. Duperrier, N. Pichoff, and D. Uriot Phys. Rev. ST Accel. Beams **10**, 084201 – Published 1 August 2007

[27] mm wave Appl. Phys. Lett. 117, 073502 (2020); https://doi.org/10.1063/5.0011397

[28] "Terahertz-driven linear electron acceleration" Emilio A. Nanni et al; ARTICLE NATURE COMMUNICATIONS, DOI: 10.1038/ncomms9486

[29] Coupling power into accelerating mode of a three-dimensional silicon woodpile photonic band-gap waveguide. Z. Wu et al; Phys. Rev. ST Accel. Beams 17, 081301 – Published 19 August 2014

[30] Cowan, B.M. "Three-dimensional dielectric photonic crystal structures for laser-driven acceleration". Phys. Rev. STAB 2008, 11, 011301.

[31] Pompili, R., Alesini, D., Anania, M.P. et al. "Energy spread minimization in a beam-driven plasma wakefield accelerator". Nat. Phys. (2021). https://doi.org/10.1038/s41567-020-01116-9

[32] Comprehensive Analytical Chemistry Volume 69 Chemical ImagingAnalysis Freddy Adams Carlo Barbante Elsevier Radarweg 29, PO Box 211, 1000 AE Amsterdam, Netherlands The Boulevard, Langford Lane, Kidlington, Oxford OX5 1GB, UK 225 Wyman Street, Waltham, MA 02451, USA

[33] "Quest for high brightness, monochromatic noble gas ion sources" V. N. Tondare, Journal of Vacuum Science & Technology A 23, 1498 (2005); doi: 10.1116/1.2101792

[34] Rahman FH, McVey S, Farkas L, Notte JA, Tan S, Livengood RH. The prospects of a subnanometer focused neon ion beam. Scanning. 2012 Mar-Apr;34(2):129-34. doi: 10.1002/sca.20268. Epub 2011 Jul 27. PMID: 21796647.

[35] B. W. Ward, John A. Notte, and N. P. Economou "Helium ion microscope: A new tool for nanoscale microscopy and metrology", Journal of Vacuum Science & Technology B 24, 2871 (2006); doi: 10.1116/1.2357967

[36] J. J. McClelland, A. V. Steele, B. Knuffman, K. A. Twedt, A. Schwarzkopf, and T. M. Wilson, Bright focused ion beam sources based on laser-cooled atoms Applied Physics Reviews 3, 011302 (2016); https://doi.org/10.1063/1.4944491

[37] Christopher A. Bower, Kristin H. Gilchrist, Jeffrey R. Piascik, Brian R. Stoner, Srividya Natarajan, Charles B. Parker, Scott D. Wolter, and Jeffrey T. Glass "On-chip electron-impact ion source using carbon nanotube field emitters", Applied Physics Letters 90, 124102 (2007); doi: 10.1063/1.2715457

[38] L. Malferrari, F. Odorici, G. P. Veronese, R. Rizzoli, D. Mascali et al. "Modification of anisotropic plasma diffusion via auxiliary electrons emitted by a carbon nanotubes-based electron gun in an electron cyclotron resonance ion source". Rev. Sci. Instrum. 83, 02A343 (2012); doi: 10.1063/1.3673634

[39] O. Kornienko, P. T. A. Reilly, W. B. Whitten, and J. M. Ramsey, ""Field-Emission Cold-Cathode EI Source for a Microscale Ion Trap Mass Spectrometer", Anal. Chem. 2000, 72, 3, 559-562 2000 https://doi.org/10.1021/ac990962s

[40] Wade L. Fite and R. T. Brackmann Collisions of Electrons with Hydrogen Atoms. I. Ionization, Phys. Rev. 112, 1141 (1958)

[41] Ammosov MV, Delon NB and Krainov VB 1986 Tunnel ionization ofcomplex atoms and atomic ions in alternating electromagnetic fields SovietPhys.—JETP64 1191–4

[42] Rogozhkin, S.V., Lukyanchuk, A.A., Raznitsyn, O.A. et al. Atom Probe Tomography Analysis of Materials using Femtosecond-Laser Assisted Evaporation. J. Synch. Investig.12, 452–459 (2018). https://doi.org/10.1134/S1027451018030175

[43] N. Liu, X. Xu, R. Pang, P. Santhana Raman, A. Khursheed, and J. A. van Kan, "Brightness measurement of an electron impact gas ion source for proton beam writing applications" Rev. Sci. Instrum. 87, 02A903 (2016); doi: 10.1063/1.4932005

[44] X. Xu, R. Pang, P. S. Raman, R. Mariappan, A. Khursheed, J. A. van Kan, Fabrication and development of high brightness nano-aperture ion source, Microelectronic Engineering, Volume 174, 2017, Pages 20-23, https://doi.org/10.1016/j.mee.2016.12.009.

[45] van Kan, J.A., Zhang, F., Chiam, S.Y. et al. Proton beam writing: a platform technology for nanowire production. Microsyst Technol 14, 1343–1348 (2008). https://doi.org/10.1007/s00542-007-0514-

[46] Jeroen A. van Kan , Rudy Pang , Tanmoy Basu "Considerations for the nano aperture ion source: Geometrical design and electrical control" Rev. Sci. Instrum. 91, 013310 (2020); https://doi.org/10.1063/1.5128657

[47] A. Cassimi*, T. Muranaka, L. Maunoury, H. Lebius, B. Manil and B.A. Huber, Multiply-charged ion nanobeams, Int. J. Nanotechnol., Vol. 5

[48] Bellucci, V. M. Biryukov, Yu. A. Chesnokov, V. Guidi, and W. Scandale, Making microbeams and nanobeams by channeling in microstructures and nanostructures PHYSICAL REVIEW SPECIAL TOPICS - ACCELERATORS AND BEAMS,VOLUME 6, 033502 (2003)

[49] Lemell, C., Burgdörfer, J. & Aumayr, F. Interaction of charged particles with insulating capillary targets – The guiding effect. Prog. Surf. Sci. 88, 237–278 (2013).

[50] Ion guiding in alumina capillaries: MCP images of the transmitted ions", Z. Juhász,*, B. Sulik, S. Biri et al;

[51] L. Cosentino, P. Finocchiaro, A. Pappalardo, A beam diagnostic multisensor for low energy radioactive beams, Nuclear Instruments and Methods in Physics Research Section A: Accelerators, Spectrometers, Detectors and Associated Equipment, Volume 622, Issue 3, 2010, 512-517, 0168-9002, https://doi.org/10.1016/j.nima.2010.07.085.

[52] J. Harasimowicz, L. Cosentino, P. Finocchiaro, A. Pappalardo, and C. P.Welsch. Scintillating screens sensitivity and resolution studies for low energy, low intensity beam diagnostics Review of Scientific Instruments 81, 103302 (2010); https://doi.org/10.1063/1.3488123

[53] G. L. Orlandi, C. David, E. Ferrari, V. A. Guzenko, R. Ischebeck, E. Prat, B. Hermann, M. Ferianis, G. Penco, M. Veronese, N. Cefarin, S. Dal Zilio, and M. Lazzarino Nanofabricated free-standing wire scanners for beam diagnostics with submicrometer resolution, Phys. Rev. Accel. Beams **23**, 042802 -2020

[54] Barberet, P., "First results obtained using the CENBG nanobeam line: Performances and applications", Nuclear Instruments and Methods in Physics Research B, vol. 269, no. 20, pp. 2163-2167, 2011. doi: 10.1016/j.nimb.2011.02.036.

[55] Ruggiero, A.G. Nuclear fusion of protons with ions of boron. Nuov Cim A 106, 1959-1963 (1993). https://doi.org/10.1007/BF02780602

[56] K. Prakrajang, J.C.G. Jeynes, M.J. Merchant, K. Kirkby, N. Kirkby, P. Thopan, L.D. Yu, MeV single-ion beam irradiation of mammalian cells using the Surrey vertical nanobeam, compared with broad proton beam and X-ray irradiations, Nuclear Instruments and Methods in Physics Research Section B: Beam Interactions with Materials and Atoms, Volume 307, 2013, pages 586-591

[57] Cirrone GAP, Cuttone G, Raffaele L, Salamone V, Avitabile T, Privitera G, Spatola C, Margarone D, Patti V, Petringa G, Romano F, Russo A, Russo A, Sabini MG, Scuderi V, Schillaci F and Valastro LM (2017) Clinical and Research Activities at the CATANA Facility of INFN-LNS: From the Conventional Hadrontherapy to the Laser-Driven Approach. Front. Oncol. 7:223. doi: 10.3389/fonc.2017.00223

[58] Ph. Barberet, S. Incerti, F. Andersson, F. Delalee, L. Serani, Ph. Moretto, Technical description of the CENBG nanobeam line. Nuclear Instruments and Methods in Physics Research B 267 (2009) 2003-2007 barberet2009

[59] S. Incerti a,*, Q. Zhang a, F. Andersson a, Ph. Moretto a, G.W. Grime b, M.J. Merchant b, D.T. Nguyen a, C. Habchi a, T. Pouthier a, H. Seznec,,Monte Carlo simulation of the CENBG microbeam and nanobeam lines with the Geant4 toolkit Nuclear Instruments and Methods in Physics Research B 260 (2007) 20-27 Incerti_2007

[60] F. Avraham Dilmanian, John G. Eley, Sunil Krishnan (2015). Minibeam Therapy With Protons and Light Ions: Physical Feasibility and Potential to Reduce Radiation Side Effects and to Facilitate Hypofractionation, International Journal of Radiation Oncology*Biology*Physics, 92, 2,2015,Pages 469-474, https://doi.org/10.1016/j.ijrobp.2015.01.018. dilmanian2015

[61] Dos Santos, M., Delorme, R., Salmon, R. and Prezado, Y. (2020), Minibeam radiation therapy: A micro- and nano-dosimetry Monte Carlo study. Med. Phys., 47: 1379-1390. https://doi.org/10.1002/mp.14009

[62] Loïc Grevillot et al. Clinical implementation and commissioning of the MedAustron Particle Therapy Accelerator for non-isocentric scanned proton beam treatments. Med Phys. 2020 Feb;47(2):380-392. doi: 10.1002/mp.13928. Epub 2019 Dec 29.

[63] The CNAO dose delivery system for modulated scanning ion beam radiotherapy, S. Giordanengo at al, https://doi.org/10.1118/1.4903276 January 2015, Medical Physics 42(1):263.

[64] Price KM and Schettino G. Microbeams in radiation biology: review and critical comparison. Radiation Protection Dosimetry (2011), 143, No. 2-4, pp. 335-339

[65] Wu LJ et al. Targeted cytoplasmic irradiation with alpha particles induces mutations in mammalian cells. Proc. Natl Acad. Sci. (1999) 96, 4959-4964.

[66] Shao C. et al. Targeted cytoplasmic irradiation induces bystander responses. Proc. Natl Acad. Sci. (2004) 101, 13495-13500.

[67] Jacob B. et al. Differential Repair Protein Recruitment at Sites of Clustered and Isolated DNA Double-Strand Breaks Produced by High-Energy Heavy Ions. Scientific Reports volume 10, Article number: 1443 (2020)

[68] Belyakov, O. V., Mitchell, S. A., Parikh, D., Randers-Pehrson, G., Marino, S. A., Amundson, S. A., Geard, C. R. and Brenner, D. J. Biological effects in unirradiated human tissue induced by radiation damage up to 1mm away. Proc. Natl Acad. Sci. (2005) 102, 14203-14208.

[69] Zhexin Zhao, Dylan S. Black, R. Joel England, Tyler W. Hughes, Yu Miao, Olav Solgaard, Robert L. Byer, and Shanhui Fan, "Design of a multichannel photonic crystal dielectric laser accelerator," Photon. Res. 8, 1586-1598 (2020)
[]U. Amaldi et al, Eur. Phys. J. Plus (2011) 126: 70 DOI 10.1140/epjp/i2011-11070-4]